# Direct ESR evidence for magnetic behaviour of graphene

S. S. Rao[*] and A. Stesmans

Department of Physics and INPAC-Institute for Nanoscale Physics and Chemistry,

University of Leuven, Celestijnenlaan 200 D, B-3001 Leuven, Belgium

Y. Wang and Y. Chen

Institute of Polymer Chemistry, Nankai University, Tianjin 300071, School of Chemistry

and Chemical Engineering, Nanjing University, Nanjing 210093, China

**Abstract**

Recently, there have appeared theoretical works on the magnetic properties of graphene and graphene nanoribbons envisaging possible spin-based applications along with fundamental scientific insight. The theoretical efforts, however, appear not paralleled by experimental investigation to test magnetic properties. Yet, room temperature ferromagnetism (RTFM) has recently been experimentally reported in graphene (G-600) [Nano. Letters **9**, 220 (2009)], the origin of which remains still unexplored. Inspired by this observation, and in attempt to trace the origin of RTFM, we here report on low temperature K-band electron spin resonance (ESR) observations on G-600. Two distinct C-related paramagnetic signals are revealed, both of a Lorentzian shape: a) a broad at g = 2.00278 which can be attributed to graphitic-like (GL) carbon; b) a narrower signal at g = 2.00288 which is associated with free radical like (FL) carbon. No other signals could be detected. We speculate that the GL ESR signal may come from the conductive $\pi$-carriers propagating in the interior of graphene sheets, while the FL ESR signal may stem from the edges of graphene sheets due to non-bonding localized electronic states. It is suggested that the long range direct/indirect exchange interaction between GL and FL C-related magnetic spin centers may lead to the reported RTFM, this pointing to C origin of the later.

**\*Corresponding author. Fax:** + 32 16 32 79 87. E-mail address:
Srinivasarao.singamaneni@fys.kuleuven.be (S S Rao)

# 1. Introduction

Carbon based materials, such as graphene, graphene oxide (GO) and graphene nanoribbons (GNRs) have recently attracted large interest both from scientific and technological point of view. In particular, graphene, a zero gap semi-metal, is characterized as "the thinnest material in our universe" exhibiting a two-dimensional (2D) honeycomb structure of carbon. It has a band structure showing two intersecting bands at two inequivalent k points in the reciprocal space. It exhibits novel electronic properties such as ballistic transport, massless Dirac fermions, Berry's phase, minimum conductivity and localization suppression [1-3]. It has been reported that graphene shows very high electron mobility ($\approx 10^5$ cm$^2$/V.S) at low temperature [4]. It promises a diverse range of applications in microelectronics, high speed optical communication devices, in spintronics, and as room temperature gas sensors [5-8].

On the other hand, magnetism in C-based materials is a rapidly evolving field of research with strong implications for spin based information technology. C-based magnetism without magnetic elements, having only s and p electrons is intriguing, particularly, when extracted from graphene, a material touted as the basis unit for future spintronic devices owing to its long spin diffusion length, small spin-orbit (SO) as well as hyperfine (hf) couplings [9]. Understanding the origin and basic mechanism behind the magnetic behaviour of C-based materials and engineering ferromagnetic (FM) carbon structures has been of prime importance since long. In their theoretical work, Yazyev *et al.* [7] have concluded that flat-band quasi localized (QL) states induced by point defects are responsible for magnetism of graphene. Lin *et al.* [10] have argued the importance of electron-electron interaction and carrier concentration in explaining the FM properties of GNRs. Theoretical calculations of Sawada *et al.* [11] indicated that zig-zag GNRs (ZGNRs) exhibit anti-ferromagnetic (AFM) phase and also argued that magnetism can be induced by carrier doping. Lehtinen *et al.* [12] have calculated that the magnetic moment due to an adatom defect is $0.5\mu_B$, $\mu_B$ being the Bohr magneton while a vacancy has a magnetic moment of about $1\mu_B$, with both defects prone to induce spontaneous long-range FM ordering. In fact, a defective graphene is predicted [13] to show room temperature ferromagnetism (RTFM). A FM phase of mixed $sp^2$ and $sp^3$ pure carbons has been predicted [14].

FM has been observed experimentally by Esquinzai *et al.* [15] on the highly oriented pyrolytic graphite (HOPG) material with a $T_c$ above 300 K when proton irradiated, and unstable (temporary) FM was observed [16] in light-weight carbon nano-foam also, − the works indicating the origin of the magnetism to be carbon related, not correlated with any external magnetic impurity. Recent experiments [17] have shown that proton irradiation doubles the magnetic moment in comparison to that observed after He ion bombardment. Recent experiments [18] employing magnetic force microscopy (MFM) and scanning tunneling microscopy (STM) techniques performed on HOPG sample revealed that 2D networks of point defects exhibiting localized electronic states are the main source of magnetism.

As described above, though there have been several experimental studies on HOPG exploring its magnetic nature, not much experimental work [19,20] has been reported to investigate the magnetic nature of graphene, a building block of HOPG. Theoretical predictions suggested that the edges and defects present in graphene enhance the DOS, which further leads to a FM phase. As predicted, the later exhibits a phase transition from the FM to paramagnetic (PM) state, that is of first order [21]. Such predictions would need experimental verification.

Wang *et al.* [22] have recently experimentally reported the RTFM in graphene (G-600) by using DC superconducting quantum interference device (SQUID) magnetization measurements. Its origin still remains unexplored. Motivated by this observation, in the present work, we first apply electron spin resonance (ESR) in an attempt to identify paramagnetic (spin) point defects potentially responsible for RTFM in G-600. We present ESR results obtained on samples (G-600), in the pristine state and after annealing in Ar at $800^{\circ}$C (G-800) or $H_2$ at $600^{\circ}$C (H-G-600). The study reveals the occurrence of two distinct, intrinsic C-related types of defects both exhibiting a Lorentzian shape. From this observation, together with the absence of any other signals, it is tentatively suggested the RTFM to be of C origin.

## 2. Experimental

Samples studied were nm-scale graphene, denoted as G-600, prepared by the modified Hummer's method [22]. Separate samples were additionally annealed at $800^{\circ}$C in Ar (1 atm) for ~ 3 h (sample G-800) or (1 atm) at $600^{\circ}$C for ~ 2 h (sample H-G-600).

Conventional K-band (~20.6 GHz) first harmonic ESR measurements were performed at low temperature (T ≤ 10 K) [23], with the spectrometer routinely operated under conditions of adiabatic slow passage. The areal spin density was quantified by the double numerical integration of the derivative absorption spectra of the computer simulated signal. Absolute spin densities were obtained making use of a co-mounted calibrated Si:P intensity marker of g(1.7 K = 1.99876). The modulation amplitude ($B_m$) of the applied magnetic field and the incident microwave power ($P_\mu$) were restricted to levels not causing (noticeable) signal distortion. No K-band measurements could be performed for T ≥10 K because of drastic deterioration of the quality factor 'Q' of the microwave cavity likely due to the strong semiconducting nature of the graphene samples.

## 3. Experimental Results and Analysis

Atomic absorption spectroscopy (AAS) indicated that G-600 does not contain significant amounts (~ 48.6 ppm) of magnetic impurities (Fe, Co, Ni). DC magnetization measurements were performed on G-600 and G-800 using the SQUID technique by sweeping the magnetic field (H) at fixed Ts of which the first results were published previously [22]. Salient features include: The magnetization (M) versus field (H) isotherms on the pristine sample (G-600) obtained at 2 and 300 K clearly show a tendency towards saturation, with a small coerceive field ($H_c$) of 40 G (at 300 K), which is a signature of FM; the hysteresis (M-H loop) associated with G-600 is feeble, as expected for a soft FM material. The M-H loop of G-600 was found to disappear upon TA at 800°C (G-800) in Ar. Wang *et al.* [22] have concluded that the observed RTFM originates from the defects present in G-600. *But, a most critical point concerns the nature of the defects causing the long-range FM coupling.* For this end, to get in sight, we have applied ESR in an attempt to uncover the magnetic species (defects) giving rise to the appearance and disappearance of FM in G-600 and G-800, and address the possible mechanisms leading to FM. In the following sections, we present the K-band ESR results obtained on G-600, G-800 and H-G-600.

## 3.1. G-600

Figure 1 shows a K-band (~ 20.6 GHz) ESR spectrum observed on G-600 at 1.7 K. Two symmetric, isotropic ESR signals are observed of zero–crossing g values $g_c$ = 2.00278 and $g_c$ = 2.00288 with corresponding spin densities $1 \times 10^{16}$ spins/g and $7 \times 10^{13}$ spins/g, respectively. These g values fall within the carbon ESR signal range (g = 2.0022 – 2.0035), and may be ascribed to C-related dangling bonds of spin S = ½ each. Despite intense signal averaging over broad field ranges under various extreme and optimized spectrometer parameter settings, no other signals could be observed. The signals could be computer simulated using Lorentzian line shapes with: a) a broad signal of peak-to-peak width $\Delta B_{PP} \approx$ 80-100 G at g = 2.00278, which can be attributed to graphitic-like (GL1) carbon; b) a narrower signal with $\Delta B_{PP} \sim$ 4.5 G, g = 2.00288, attributed to free radical like (FL1) carbon. The fact that no other signals could be detected suggest that the RTFM is of C origin. We speculate that the GL1 ESR signal may come from the mobile $\pi$-carriers propagating in the interiors of graphene sheets, while the FL1 ESR signal may stem from non-bonding localized states at the edges of graphene sheets. Similar assignment has been made earlier [24] for the ESR signal originating from graphitic like pyrocarbon. To probe deeper, Q-band ESR spectroscopy was attempted, yet without success (even at low temperatures) because of resonator problems.

### 3.2. G-800

Wang *et al.* [22] have shown that FM has disappeared in G-800, which in view of the results obtained on G-600, would make ESR analysis of much interest. Accordingly, K-band ESR measurements have been carried out on G-800 down to 1.8 K. The obtained spectrum is shown in Fig 2. Measurements at higher temperatures ( $\geq$ 4.2 K) were drastically hampered due to excessive cavity loading. As illustrated by the ESR spectrum observed at 1.8 K, no ESR signal could be detected originating from G-800, either intrinsic or extrinsic in nature. This result suggests that the disappearance of FM after TA at $800^{o}$ C, may be linked to elimination of ESR active centers [22].

### 3.3. H-G-600

      To further assess the nature of the above revealed defects (GL1 and FL1), in a next step, G-600 was treated in $H_2$ at $600^o$ C, Fig. 3 showing a K-band ESR spectrum (upper curve) measured on H-G-600. Though the observed signals weakened, we keep observing two Lorentzian shaped signals. The estimated spin densities of GL2 and FL2 are $2 \times 10^{14}$ and $3 \times 10^{12}$ spins/g, respectively through computer simulations (bottom curve). Yet, both signals, now labeled GL2 and FL2, appear broadened by a factor $\approx 2$ and $\approx 3$, respectively, compared to the respective signals observed in G-600; ($\Delta B_{PP}$(GL2) = 160-170 G, $\Delta B_{PP}$(FL2) = 15 G), with inferred g values $g_c$ (GL2) = 2.0028 and $g_c$(FL2) = 2.00296, that is, unchanged within the experimental accuracy ($\pm$ 0.0003). Noteworthy is that in the case of H-G-600, the loading of the cavity by the sample has been drastically reduced, indicating that the electronic properties of G-600 are much modified upon $H_2$ treatment.

## 4. Discussion

      We speculate that the GL ESR signal arises from the conductive $\pi$-carriers propagating in the interiors of graphene sheets, while the FL ESR signal may stem from non-bonding localized electronic states at the edges of graphene sheets. We further propose that the long range direct/indirect exchange interaction between GL and FL C-related magnetic spin centers may lead to the observed RTFM. Previously, various forms of hydrogen-adsorbed defects have been suggested as possible origin of localized magnetic moments. The current results, with no obvious presence of hyperfine (hf) structure, can be taken as negative evidence for that suggestion.

The findings of Wang and co-authors [22] are well corroborated by our current ESR results in terms of the appearance and disappearance of ESR signals in G-600 and G-800 respectively. The ESR data would disqualify the extrinsic metallic impurities as the possible origin of RTFM. From the ESR results, we also infer that the proposed defect is not a di-vacancy, as no forbidden ESR transition, a characteristic feature of spin S=1 centers, could be traced.

The question of magnetic ordering in defective graphene at finite temperatures remains largely unaddressed. Regarding the mechanism for the formation of the FM state in G-600, among others, the defect mediated mechanism appears to be the favored one. The single atom and localized defect mediated mechanism has been addressed in a number of publications. The current results, reporting the presence of two paramagnetic centers, may suggest such interpretation, meriting some further discussion as to the possible origin of FM in G-600: (i) as known, graphene is a bipartite lattice with two interpenetrating sub-lattices. According to Lieb's theorem [25], a single vacancy (defect) results in the formation of a local magnetic moment with S = 1/2. From the theoretical calculations, it is known that magnetic moments pertaining to the same sublattice couple ferro-magnetically while moments pertaining to different sub-lattices couple anti-ferromagnetically. Clearly, the later case does not apply here. Based on the fact that FM is observed in G-600, one may argue that the magnetic moments present in the same sub-lattice would give rise to such a phenomena. (ii) Next, it may also be proposed that the RKKY-type interaction [26] among the magnetic moments via conduction carriers present in the same sub-lattice can also lead to FM features. (iii) It may also be possible that the interaction between $sp^2$ and $sp^3$ type spin units can give rise to the reported FM [14].(iv) Finally, it has also been argued that the negative Gaussian curvatures in graphene-layers [27] and weak magnetic coupling between the individual layers may lead to the formation of localized magnetic moments [13].

According to a study [28], based on density functional theory, of magnetism in proton-irradiated graphite, it has been concluded that the H-vacancy complex plays a predominant role in the magnetic behaviour. Yet, as mentioned, we failed to find any indication for such hydrogen-vacancy complexes from ESR finger print signatures. Also, should there be unobserved unpaired electron involved on an edge oxygen, an anisotropic g value is expected which would reflect in the observation of powder pattern line shapes, unlike observations.

## 5. Summery and Conclusions

Low temperature electron spin resonance measurements have revealed the presence of two distinct paramagnetic C-related spin centers in graphene prepared by the modified Hummer's method, which were found to be eliminated after thermal annealing at 800 $^{o}$C. Among the suggested possible scenarios for the onset of ferromagnetism in graphene-related materials, we believe that the observed RTFM may originate either from the direct/indirect long range exchange interaction between these two spin centers or through an RKKY type interaction between spin centers pertaining to the same C-sublattice. Our ESR experimental results lead us to reject FM impurities as the origin of the observed magnetism. ESR experiments in combination with thermal treatment (Ar, $H_2$) indicate that the possibility for tuning the magnetization in graphene by thermal steps. A dramatic change in conductivity is noticed upon treatment in $H_2$. Further work is in progress to probe the nature of spin dynamics of the revealed defect centers using X-band spectroscopy.

# References

[1] A. K. Geim and K. S. Novoselov, Nat. Mater. 6 (2007) 183.

[2] A. H. C. Neto, F. Guinea, N. M. R. Peres, K. S. Novoselov and A. K. Geim, Rev.

   Mod. Phys. 81(2009) 109.

[3] Y. Zhang, J. W Tan, H. L. Stormer and P. Kim, Nature 438 (2005) 201.

[4] K. I. Bolotin, K. J. Sikes, Z. Jiang, M. Klima, G. Fudenberg, J. Hone, P. Kim and

   H. L. Stormer, Solid State Commun. 146 (2008) 351.

[5] P. Avouris, Z. Chen and V.Perebeinos. Nature Nanotech. 2 (2007) 605.

[6] T. Mueller, F. Xia and P. Avouris, Nature Photonics 4 (2010) 297.

[7] O. V. Yazyev and M. I. Katsnelson, Phys. Rev. Lett. 100 (2008) 047209.

[8] F. Schedin, A. K. Geim, S. V. morozov, E. W. Hill, P. Blake, M. I.

   Katsnelson and K. S. Novoselov, Nature Materials, 6 (2007) 652.

[9] N. Tombros, S. Tanabe, A. Veligura, C. Jozsa, M. Popinciuc, H. T. Jonkman and B. J.
   van Wees, Phys. Rev. Lett. 101 (2008) 046601.

[10] H. H. Lin, T.Hikihara, H-T. Jeng, B-L Huang, C-Y Mou and X. Hu, Phys. Rev. B.
   79 (2009) 035405.

[11] K. Sawada, F. Ishii, M. Saito, S. Okada and T. Kawai, Nano Lett. 9 (2009) 269.

[12] P.O. Lehtinen, A. S. Foster, A. Ayuela, A. Krasheninnikov, K. Nordlund and R.M.

   Nieminen, Phys. Rev. Lett. 91 (2003) 017202.

[13] L. Pisani, B. Montanari and N. M. Harrison, New J. Phys. 10 (2008) 033002.

[14] D. Arčon , Z. Jagličič , A. Zorko , A. V. Rode , A. G. Christy , N. R. Madsen, E. G.

   Gamaly and B. L. Davies, Phys. Rev. B 74 (2006) 014438, and references there in.

[15] P. Esquinazi, D. Spemann, R. Hohne, A. Setzer, K.-H. Han, T. Butz,

   Phys.Rev. Lett. 91 (2003) 227201.

[16] A. V. Rode, E. G. Gamaly, A. G. Christy, J. G. F. Gerald, S. T. Hyde, R. G. Elliman,
B. L-Davies, A. I. Veinger, J. Androulakis and J. Giapintzakis, Phys. Rev. B 70
(2004) 054407

[17] D. Spemann, K.-H. Han, P. Esquinazi, R . Hohne and T. Butz, Nuc. Inst. and Meth.
in Phys. Res. B 219 (2004) 886.

[18] J. Cervenka, M. I. Katsnelson and C. F. J. Flipse, Nature Physics 5 (2009) 840.

[19] H. S. S. Ramakrishna Matte, K. S. Subrahmanyam and C. N. R. Rao, J. Phys. Chem.
C 113 (2009) 9982.

[20] Luka Ćirić [*], Andrzej Sienkiewicz [*], Bálint Náfrádi, Marijana Mionić, Arnaud
Magrez, László Forró, Phys. Status Solidi B 246 (2009) 2558.

[21] E. V. Castro, N. M. R. Peres, T. Stauber and N. A. P. Silva, Phys. Rev. Lett. 100
(2008) 186803

[22] Y. Wang, Y Huang, Y. Song, X. Zhang, Y. Ma, J. Liang and Y. Chen, Nano. Lett. 9
(2009) 220.

[23] A. Stesmans. Phys. Rev. B 48 (1993) 2418.

[24] C. Buschhaus and E. Dormann, Phys. Rev. B 66 (2002) 195401.

[25] E.H. Lieb, Phys. Rev. Lett. 62 (1989)1201.

[26] M. A. H.Vozmediano, M.P. L-Sancho, T. Stauber, F. Guinea, Phys. Rev. B
72 (2005)155121.

[27] N. Park, M. Yoon, S. Berber, J. Ihm, E. Osawa, and D. Tom´anek, Phys. Rev. Lett.
91 (2003) 237204.

[28] P. O. Lehtinen, A. S. Foster, Y. Ma, A. V. Krasheninnikov and R. M. Nieminen,
Phys. Rev. Lett. 93 (2004) 187202.

**Figure Captions**

**FIG. 1.** Derivative-absorption K-band ESR spectrum (~ 50 cumulative scans) observed at 1.7 K on G-600 using $B_m$ = 0.5 G and $P_\mu \approx$ 2.5 nW. The dashed curve (bottom) represents a computer simulation (combined) of the observed signals at $g_c$ (GL1) = 2.00278 and $g_c$ (FL1) = 2.00288 of the corresponding peak-to-peak line widths $\Delta B_{PP}$ (GL1) = 80-100 G and $\Delta B_{PP}$ (FL1) = 4.5 G. The narrow signal at $g_c$ = 1.99876 stems from a co-mounted Si:P marker sample. Poor signal-to-noise ratio is due to strong loading the Q factor of the cavity.

**FIG. 2.** K-band ESR spectrum measured on G-800 at 1.8 K ($B_m$ = 0.5 G and $P_\mu$ = 2.5 nW). There is no indication of a C-originated ESR signal.

**FIG. 3.** K-band ESR spectrum (top curve) observed at 1.7 K on H-G-600 using $B_m$ = 0.5 G and $P_\mu$ = 2.5 nW. The small low field signal might arise from noise. The dashed curve (bottom) represents a computer simulation of the observed signal at $g_c$ (GL2) = 2.0028 and $g_c$ (FL2) = 2.00296.

**Figures**

**FIG. 1.**

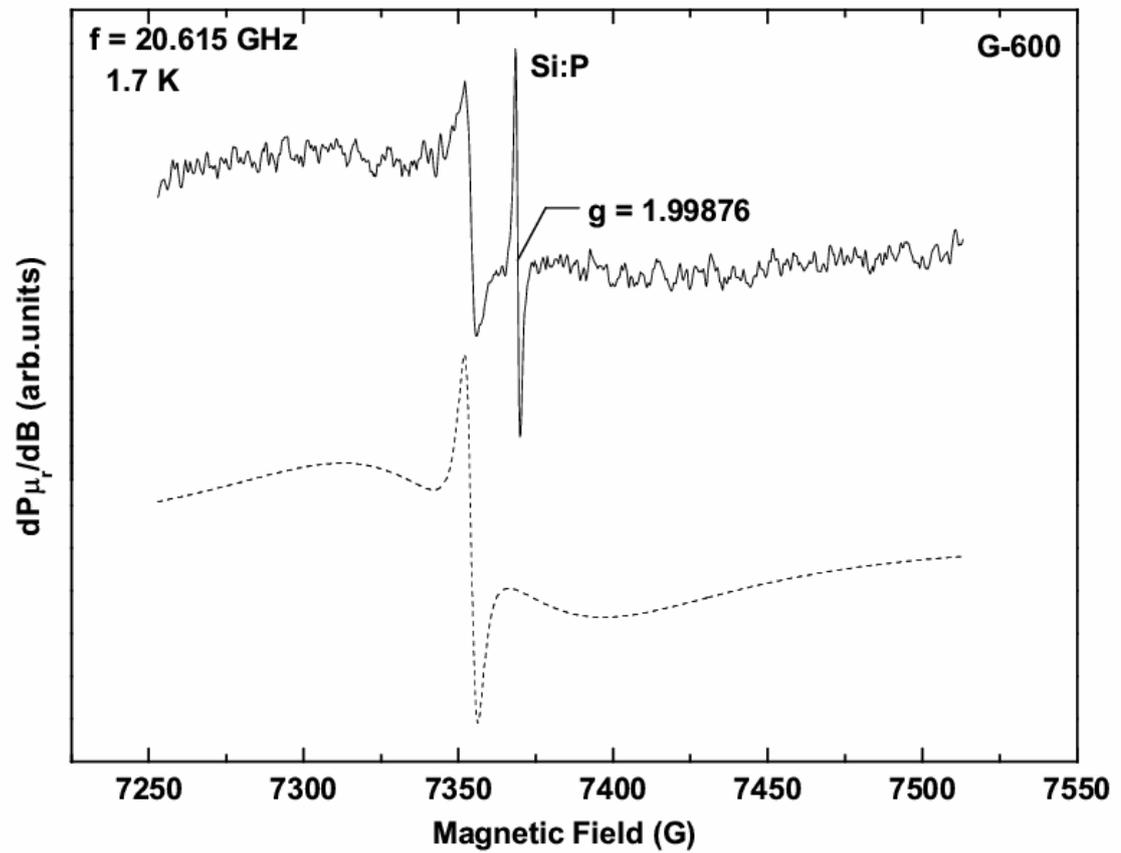

**FIG. 2.**

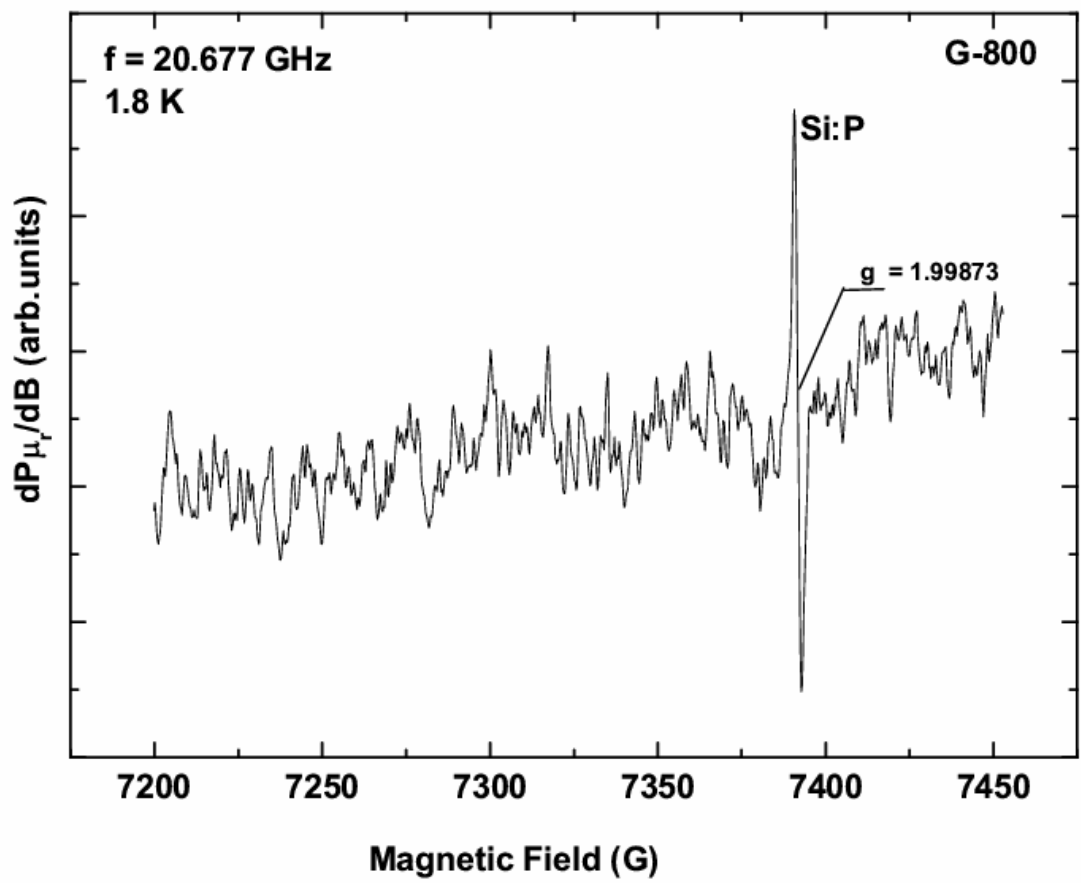

**FIG. 3.**

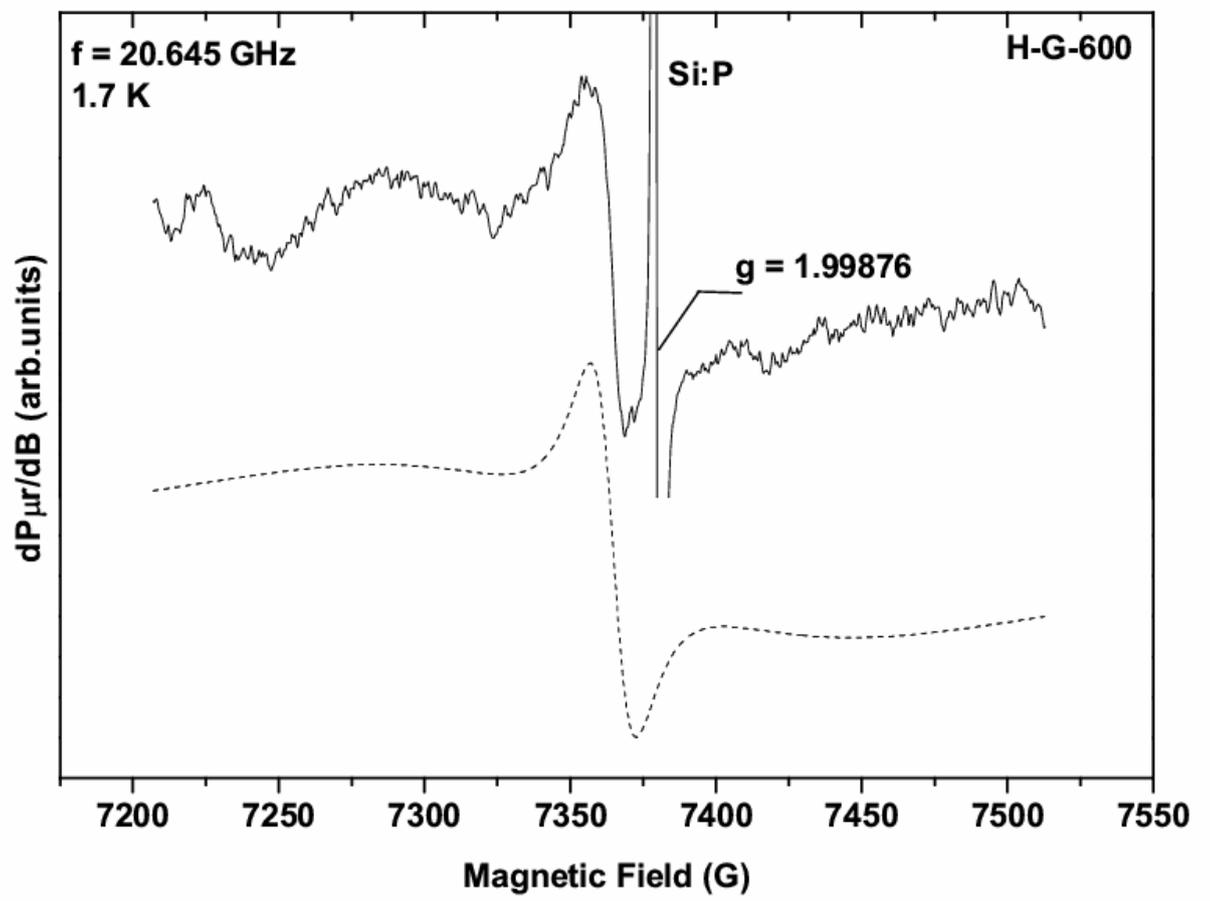